\documentclass[a4paper]{jpconf}
\usepackage{graphicx}
\usepackage[utf8x]{inputenc}
\usepackage{wrapfig}
\usepackage{float}
\usepackage{amsmath}
\begin{document}
\vspace*{-1in}
\title{Appraisal of the realistic accuracy of Molecular dynamics of high pressure hydrogen}
\author{Graeme J Ackland and Ioan B Magdău}
\address{CSEC, SUPA, School of Physics and Astronomy, The University of Edinburgh, Edinburgh EH9 3JZ, United Kingdom}
\ead{gjackland@ed.ac.uk, i.b.magdau@sms.ed.ac.uk}
\begin{abstract}

Molecular dynamics is a powerful method for studying the behaviour of
materials at high temperature.  In practice, however, its
effectiveness in representing real systems is limited by the accuracy
of the forces, finite size effects, quantization, and equilibration
methods.  In this paper we report and discuss some calculations
carried out using molecular dynamics on high pressure hydrogen,
reviewing a number of sources of error. We find that the neglect of
zero-point vibrations is quantitively the largest.  We show that simulations
using ab initio molecular dynamics with the PBE functional predict a
large stability field for the molecular Cmca-4 structure at pressures
just above those achieved in current experiments above the stability
range of the mixed- molecular layered Phase IV.  However, the various
errors in the simulation all point towards a much smaller stability
range, and the likelihood of a non-molecular phase based on
low-coordination networks or chains of atoms.
\end{abstract}

\section{Introduction}

The theory of molecular dynamics (MD) is well established and MD
remains the best way to study high temperature behaviour of materials,
especially dynamical processes such as phase transitions, anharmonic
vibrations and diffusion.
With classical molecular dynamics, the number of atoms which can be
routinely treated is now well above 10$^7$.  As a consequence many of
the finite size effects which early protagonists wrestled with are
eliminated.  However, with the advent of ab initio molecular dynamics (AIMD),
system sizes have dropped back to 10$^2$, and the problems met by
classical MD in the 1980s have reemerged.

As long ago as 1935, Wigner and Huntingdon\cite{wigner1935possibility} predicted that with a
tenfold increase in density, metallic hydrogen might be formed at high
pressure.  Evidence for metallic hydrogen comes from the strong
magnetic fields of gas giant planets\cite{Planets,Jupiter}.  On earth,
such pressures are accessible at high-temperature in shock-wave
experiments\cite{Fortov,Nellis}.  There are also at least three
distinct liquid structures \cite{Ashcroft2000}.
Compression using diamond anvil cells in now approaching the metallization density 
predicted by Wigner and Huntingdon\cite{wigner1935possibility}, and the general structure of
the phase diagram of hydrogen is becoming established, with
tantalising similarity to other mono- and divalent
elements\cite{Heine,RbChain,ReedBa2000,marques2009potassium,
  ma2009transparent,marques2011sodium,arapanaSc,marques2011crystal}. At
low pressure ``Phase I'' adopts a hexagonal close packed structure of
freely rotating (i.e. spherical on average) molecules.  As pressure
increases the rotation becomes more correlated between molecules, and
at low temperature a broken symmetry phase freezes in (Phase II).  At
still higher pressure electrons begin to delocalize forming molecular
trimer units (Phase III), while at higher T a mixed structure of
trimers and hindered rotors appears (Phase IV).  At these pressures 
the liquid is denser than the solid, so the melt curve has
negative slope.

All these qualitative features are captured by AIMD, even using a few
hundred atoms.  At still higher pressures it is expected that a
transition from molecular to atomic phases will occur. From static
calculations, candidate phases above 300GPa are molecular $Cmca$ or
atomic $I4/amd$ (Cs-IV)
phases\cite{pickard2007structure,pickard2012density,geng2012high}.
However, previous experience in divalent metals suggests that very
complicated structures may exist, beyond what can be found in
structure searches.  Phase IV shows that the high temperature phase
may be very different from the low-T phase.

The classical treatment of nuclei brings a raft of problems which are
particularly significant for hydrogen.  In particular, energy in
molecular dynamics is divided between classical degrees of freedom
according to equipartition, while in quantum reality the degrees of
freedom should be populated according to quantum statistics. 

Here, we compare and contrast AIMD and two methods for including
quantum protons: path integral molecular dynamics (PIMD) which
quantizes motion in real space and lattice dynamics which takes phonon
occupancy as the good quantum number. We also examine different exchange-correlation
functionals to determine the most significant sources of uncertainty
in current theory.

\section{Calculation Details}
\subsection{Ab initio molecular dynamics}
Our molecular dynamics proceed based on the forces between atoms in the NPH ensemble.  In
this work we generate forces using density functional theory and
Kohn-Sham Hamiltonian \cite{kohn1965self}
implemented in the CASTEP program \cite{segall2002first,CASTEP}, with norm conserving
pseudopotentials and various exchange correlation functionals \cite{perdew1996generalized,perdew2008restoring,ceperley1980ground, perdew1981self}.

\begin{wrapfigure}[24]{R}{80mm}
  \includegraphics[width=80mm]{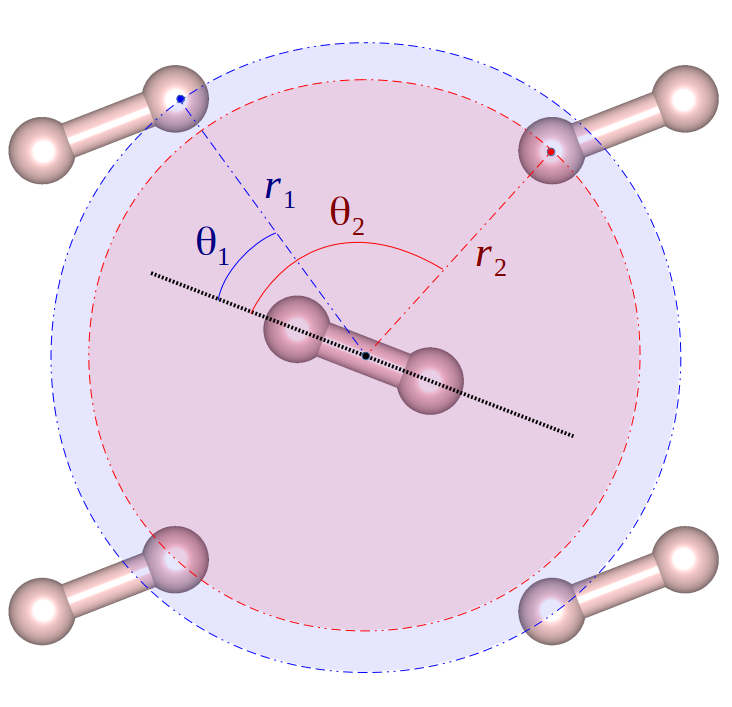}
  \caption{Definition of angular distribution function (ADF). Cmca-4 is molecular so we bin up the angles defined by the axis of a reference molecule and neighbour atoms. The histogram is weighted with the fourth power of the distance $r$ such that the further the neighbours the less they contribute to the distribution.\label{adf}}
\end{wrapfigure}

We calculate radial distribution functions (RDF) and angular
distribution functions (ADF) (see figure \ref{adf}), while our
dynamical calculation of phonon density of states (PDOS) is done with
a sliding window technique over a simulation of length T (after
equilibration) to eliminate dependence on the initial phase of the
vibration:
\begin{gather}
g(\omega) \propto \int_0^T\int_0^T\sum_{j=1}^Ne^{iwt}v_j(t+\tau)v_j(t)d\tau dt
\end{gather}
where $v_j(t)$ is the velocity of atom $j$ at time $t$ in the
simulation.  This density of states can also be projected onto
individual modes, following methods developed by Pinsook
\cite{pinsook1999calculation}, to extract specific details from a
simulation such as Raman spectra \cite{magduau2013identification} or
transition paths. Our simulations are on a small 128 atom cell, which
will be subject to finite size effects, in particular diffusion can be
confounded by phase fluctuations.  In this case, however, the MSD
saturates (Fig \ref{rdf}) and the cell remains stable, indicating that we
remain within a single phase.  In the absence of any interesting
dynamical processes, we contrast 
 thermodynamic properties from these MD simulations with lattice dynamics.

\subsection{PIMD calculations}
We also carried out PIMD calculations, as implemented in CASTEP.
The essence of PIMD \cite{marx1994ab, tuckerman1993efficient}  is to
treat the nuclei as a set of distinguishable particles and use
configurations generated from MD to sample their wavefunction via the
Feynman path integral approach.  To do this sampling, one runs several
replicas of the system (known as ``beads'') in conventional molecular
dynamics, with the Hellman-Feynman forces augmented by interactions
with the ``same'' atom in other beads.   PIMD enables the
quantum energy, including zero point, to be evaluated directly at
finite temperature.
Formally, the atoms are moving in ``imaginary time'' and sampling 
a distribution, such
that their trajectories have no physical meaning\footnote{ ``Path Integral Molecular Dynamics'' is something of a
  misnomer. There are no dynamics, rather the MD technique is used to
  sample the energy landscape in Wick-rotated imaginary time.  There
  are no molecules: atoms are treated as independent, distinguishable
  entities, and the integral is replaced by a sum over a typically
  small number of replicas.}.
Nevertheless, the application of PIMD uses standard MD packages which
have atoms moving on trajectories, and it is
tempting to assign meaning to these trajectories. 
A technique known as ``centroid dynamics'' \cite{cao1994formulation} assigns physical
meaning to the motion of the atomic position averaged over all beads
with the ``imaginary time'' treated as real.  For a PIMD run with only one bead,
centroid dynamics becomes identical to a classical simulation, giving
some weight to the idea that ``centroid dynamics'' may resemble time-dependent behaviour
behaviour.
\footnote{Centroid dynamics is a marked improvement
on standard MD, however for hydrogen molecules the assumptions break
down.  For example, the hydrogen atoms in a molecule cannot be treated as
distinguishable (in classical MD, these problems become
manifest if the molecule rotates through 180 degrees).
}

\subsection{Static relaxation and Lattice dynamics}
MD follows classical trajectories, yet the actual vibrational motion
should be described by quantum mechanics.  Lattice dynamics and the harmonic approximation \cite{born1966dynamical}
provides a simple approach in which the classical vibrational
frequencies, extracted by lattice dynamics or fourier transform of the velocity
autocorrelation function (eq.1), are then treated as 3N-3 quantum harmonic oscillators.

The internal energy (kinetic and potential energy above the
ground state) of a classical harmonic crystal, assuming equipartition of energy, is:
\begin{gather}
U = (3N-3)k_BT
\end{gather}
whereas the quantized energy of harmonic oscillators is
\begin{gather}
U = \sum^{3N-3}_{i=1}  \frac{\hbar \omega_i}{\exp({\hbar \omega_i}/kT) -1} 
+ \frac{\hbar \omega_i}{2}
\end{gather}

At high temperature, these differ only by the zero point energy, so it is helpful to
include zero point energy in the ground state potential energy.  

\begin{gather}
U_q =  (3N-3) \int_{\omega=0}^{\infty} \frac{\hbar\omega g(\omega)}{\exp({\hbar \omega}/kT) -1} 
+ \frac{\hbar\omega g(\omega)}{2} d\omega
\end{gather}

Conceptually, the phonon density of states, $g(\omega)$ is
 the same classically or quantum mechanically, and can be extracted
from AIMD (Eq. 1), even for the liquid phase.
Thus it should be possible to include quantum nuclear effects by sampling the
classical normal modes.
Once we have $g(\omega)$, the quantum-correction to the internal energy from a
simulation can be readily calculated: $ \Delta U = - (3N-3)k_BT + U_q$ (see figure \ref{quasi}). 
Other thermodynamics properties also follow, e.g. the heat capacity
\begin{gather}
C_v(T) = 3k_B
\int_0^\infty \left ( \frac{\hbar\omega}{k_BT}\right )^2
\frac{\exp(\frac{\hbar\omega}{k_BT})}{
    \exp(\frac{\hbar\omega}{k_BT})-1} g(\omega)d\omega
\end{gather}    
For
hydrogen, the classical limit is well above the melting temperature
(e.g. for vibrons $\hbar\omega/k_b \approx 6000K$).
Consequently the vibrational entropy in AIMD is an order of magnitude higher than
in LD.   Furthermore, for hydrogen, zero point energy is also significant and
phase-dependent, and neglecting it will affect phase stability and 
direct calculation of
melting point, e.g. via Z-method \cite{belonoshko2006melting} or
coexistence \cite{morris1994melting}.

At the high-N limit, the distribution of frequencies becomes a
continuous density of states $g(\omega)$ and the quantum free energy becomes:

\subsection{Validity of Exchange-correlation function}
The exchange-correlation functional is the main uncontrolled approximation in density functional theory.
Probably the most reliable description of electronic exchange-correlation
is Quantum Monte Carlo (QMC)\cite{azadi2013fate,azadi2014dissociation}.  However, due to the
computational cost this is currently not usable in molecular dynamics.
For AIMD there are many different parameterized exchange-correlation
functionals in use.

The popular PBE functional has become a de facto standard for AIMD
hydrogen calculations, despite evidence that it overbinds molecular states compared with QMC\cite{azadi2013fate,azadi2014dissociation}.  
Table \ref{XC} gives a detailed comparison of
PBE\cite{perdew1996generalized} with two similar
functionals, PBESOL, which has the same form as PBE but is fitted to a
database of solids\cite{perdew2008restoring}, and the local density
approximation \cite{ceperley1980ground, perdew1981self} (LDA) fitted
to the free electron gas.  We compare the fourfold-coordinated $I41/amd$
structure with two variants of the molecular $Cmca4$ structure, differing in lattice parameter
and internal coordinate, as taken from previous work
\cite{pickard2007structure,pickard2012density}.  We tabulate harmonic
calculations corresponding to static relaxation at nominal
pressures\footnote{i.e. the volume differential of the Kohn Sham
  energy in the Born Oppenheimer approximation} of 300 and 400GPa.
As discussed later zero point and thermal pressures should be
added to these, so they are overestimates by about 50GPa. 
The most notable feature of the table is that PBE strongly favours the
molecular phases, and gives lower densities for all phases.  PBE disagrees with
more accurate QMC calculation\cite{azadi2013fate,azadi2014dissociation}, a failing which PBESOL was designed to counter.

\subsection{Simulation Results}

To contrast the methods, we have run calculations initiated in
$Cmca4$, and LD calculation contrasting this molecular phase with the
atomic $I4/amd$ structure (Table 1).  

The LD shows that, independent of exchange-correlation functional, the
$Cmca4$ phase has lower Kohn-Sham energy and much higher zero point
energy.  It is interesting that energy minimization finds two
metastable variants of Cmca (except with PBE at 300GPa): molecular
dynamics can provide a good test as to whether the they should be
regarded as part of the same ``phase'' at high temperature.
At 400GPa, $Cmca4$(low) should should be stable in AIMD and unstable in PIMD,
however no transition or phase fluctuation is observed in either case,
presumably due to kinetic barriers.  

Three RDFs and ADFs are shown:
classical AIMD, PIMD-centroid dynamics, and RDFs direct from a single
PIMD bead (Figures 2-4).
The most obvious effect is that the individual beads have an
unphysical broad spread of position.  Closer inspection shows that the
PIMD centroids also have lower, broader peaks than AIMD, due to the combination of
thermal and quantum uncertainty.  The mean squared displacement for
the three simulations have zero slope at long times, showing there is 
no diffusion or phase fluctuation.
Curiously, the long(imaginary)time atomic MSD is smaller under centroid dynamics despite 
the wider spread of bondlength.

Our NPT ensemble allows us to compare thermodynamics properties under
the two approximations (Fig 4) under various pressures.  The ZPE
shift is clear in the figure, and the associated zero point pressure means that the
``PIMD'' material is more difficult to compress, and the effect is
significant: the PIMD volume is higher by some 6\%.

  The equilibrium cell volumes in MD
  contrast with the 0K value of 163$\AA^3$, implying a thermal
  expansivity of $3\times 10^{-5}$K$^{-1}$, typical for a metallic
  solid.  From the pressure-volume relation, we can estimate the bulk modulus to
  be of order 10$^{12}$Pa, much higher than any material at ambient
  conditions.

\begin{table}[h]
\centering
\begin{tabular}{|l|l|ccc|c|}
\hline
Phase & XCF & V/atom & $\Delta$ H$_{KS}$ (eV/atom) &  ZPE (eV/atom) & v,w \\
\hline
400GPa\\
\hline
Cmca-4  & LDA & 1.271 &  -0.003 & 0.296& 0.366   0.435 \\ 
Cmca-4 (high) & LDA & 1.258& -0.003 & 0.286&  0.117   0.129 \\ 
I41-amd  & LDA &  1.237& 0 & 0.227 &\\ 
\hline
Cmca-4  & PBESOL & 1.271 & -0.004 & 0.296 & 0.366   0.436 \\ 
Cmca-4 (high)  & PBESOL  & 1.257& -0.003 & 0.286 & 0.117 0.129 \\ 
I41-amd  &  PBESOL & 1.228 & 0  & 0.226 &\\ 
\hline
Cmca-4  &PBE  & 1.280 & -0.023 & 0.301& 0.368   0.439  \\ 
Cmca-4 (high) &PBE  & 1.267 & -0.014 & 0.287& 0.111   0.131\\ 
I41-amd  & PBE & 1.245 & 0 & 0.232 &\\ 
\hline
300GPa\\
\hline
Cmca-4  & LDA & 1.424 &  -0.045 & 0.280 &  0.368   0.441  \\ 
Cmca-4 (high) & LDA & 1.408 &  -0.003 & 0.270 &  0.108   0.133  \\ 
I41-amd  & LDA &  1.411& 0 & 0.214&\\ 
\hline
Cmca-4  & PBESOL & 1.424 & -0.048 & 0.281 & 0.368   0.441\\ 
Cmca-4 (high)  & PBESOL  & 1.407 & -0.003 & 0.269 &  0.108 0.132\\ 
I41-amd  &  PBESOL & 1.410 & 0  & 0.211&\\ 
\hline
Cmca-4  &PBE  & 1.445 & -0.058 & 0.290 &  0.371   0.441 \\ 
Cmca-4  &PBE  & 1.445 & -0.058 & 0.290 &  0.133   0.059 \\ 
I41-amd  & PBE & 1.417 & 0 & 0.214&\\ 
\hline
\end{tabular}
\caption{Effect of different exchange correlation functionals on
  volume, Kohn-Sham enthalpy relative to $I41-amd$, zero-point energy.
  The three functionals are PBE, PBESOL and LDA, calculations are done
  based on static relaxations in the harmonic approximation. 
  The relaxed fractional coordinates of the two Cmca4 structures are shown: at 300GPa
  the Cmca-4 (high)\cite{pickard2007structure} structure transformed into 
the Cmca-4 structure ($u,v\rightarrow \frac{1}{2}-u,  \frac{1}{2}-v$.)
\label{XC}}
\end{table}

\section{Sources of error in MD}

\subsection{Definition of a phase}

The definition of a phase within molecular dynamics is not
straightforward.  The simulation traverses a series of microstates,
and in order to define the free energy of a phase one needs to define
which microstate belongs to which phase.  Symmetry is of marginal use,
since the microstates have no symmetry. If a simulation remains in a
single phase, time-averaged quantities can be used to define symmetry,
but where there are two distinct ``phases'' with the same symmetry
(e.g $Cmca4$).  If the simulation changes phase, or undergoes an
excursion between phases, this becomes even more problematic. A
distinct change in a continuous variable such as pressure or volume
over a short time period may define a discontinuous transformation.

\subsection{Finite size effect and phase fluctuations}

When sampling the phase space, the probability of being in a
particular phase is exponentially dependent on the number of atoms.
It means that in the thermodynamic limit phase transformations are
sharp.  However, for a finite size system, the statistical weight of
the unstable phase may be significant, even quite far from the phase
boundary.  Consider for example a typical simulation of hydrogen in Phase
IV \cite{magdau2014high} which should have layer stacking BG'BG''.  At 600K with a free
energy difference of 5 meV/molecule between stable and metastable
stacking (e.g. BBGB), simulation with 4 layers and 144 atoms could be
expected to be in the ``wrong'' phase 15\% of the time. If there is
no significant barrier the dynamics of such a simulation will be
dominated by such unphysical transitions, giving unphysical effects like 
divergent MSD\cite{magdau2014high,magduau2013identification,goncharov2013bonding}.

It is also
possible to have multiple representations of the same phase, so for
example a BG'BG'' stacking could
be represented as any of four permutations: (e.g  G''BG'B).  A
simulation which is ``stuck'' in a single representation will sample
only 1/4th of the phase space compared to one sampling the entire
space.  This will result in lower entropy $k_B\ln(4)$: negligible in the
thermodynamic limit, but in a simulation with 50 atoms at 600K, even this
is equivalent to 
a few tenths of meV per atom.

Phase fluctuation problems arise whenever it is kinetically possible for the simulation
to transform between phases, either homogeneously or by nucleation of
a defect which passes through the entire system.  Although this is
system dependent, the signature will be a short-lived rearrangement 
of all the atoms in the system, followed by a period of relative stasis.
We have observed this behaviour in some simulations which are therefore of 
no use in calculating thermodynamic properties and not reported here.

To obtain a crystal structure in MD, it is necessary to have a number
of atoms compatible with the unit cell.  This is especially
challenging when the unit cell is not known {\it a priori}.  In
hydrogen, even numbers of atoms are always used to allow molecules to
form and typically ones with many factors. However, it may still be
impossible to obtain complex phases, e.g. there are no MD studies in
hydrogen compatible with structures such as the 40 and 88 atom unit
cells in lithium\cite{marques2011crystal}, the element adjacent to H in
the Periodic Table.

Similarly, simulations run in NVE ensemble are restricted to cells
compatible with the box.  For anything other than cubic systems, such
simulations are inevitably run at inhomogeneous pressure, and any
phase transition is strongly inhibited, at best requiring
twinning\cite{Pinsook2000,Kastner2009109}.  The NPT
ensemble\cite{martovnak2003predicting,parrinello1981polymorphic}.
alleviates this somewhat, but unless some bespoke damping method is
used it suffers from a long lived ``ringing mode''.  

This is due to
poor coupling between the phonon with the same wavelength as the box
and other modes, it
extends the equilibration period and may affect the dynamics; in NVE
the same preequiibration ringing effect can be seen in the pressure.

\subsection{Classical treatment of the nuclei}
 Quantization
has a significant effect on energy.
AIMD makes a severe approximation treating the nuclei as classical
particles.  PIMD treats them as distinguishable quantum
particles. Lattice dynamics quantizes the vibrations. 

\begin{wrapfigure}[37]{L}{85mm}
  \includegraphics[width=80mm]{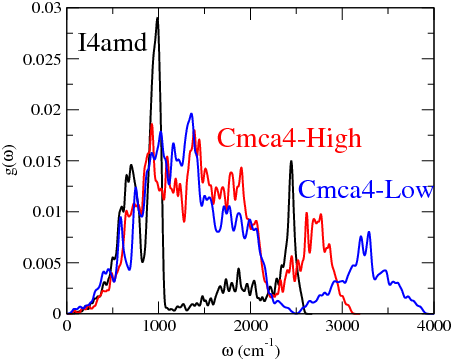}
  \includegraphics[width=80mm]{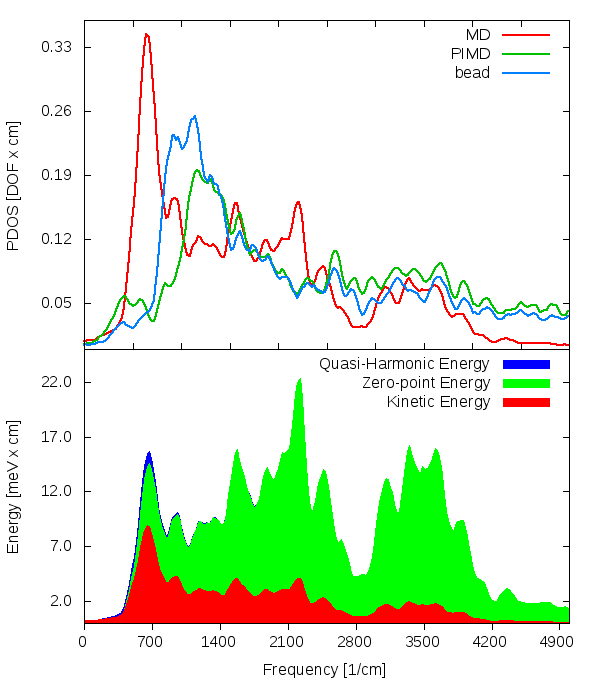}
\caption{Upper figure,
  $g(\omega)$ (PDOS) for I4amd and two Cmca-4 structures from PBESOL harmonic
  lattice dynamics at a nominal pressure of 400GPa.  Lower figure, top
  panel shows Cmca-4 PDOS extracted from 128-atom (i.e. a supercell with 8 
 layers of 8 molecules) , 400GPa, 300K MD,
  PIMD centroid dynamics and one PIMD bead using eq.1.  The curves are
  normalized to the degrees of freedom of the system. bottom panel,
  energy  distribution between the modes in MD
  (red), the zero point energy obtained by quantizing the molecular
  dynamics PDOS (green) and the total quasi-harmonic energy (zero
  point and thermal) (blue). 
\label{quasi}
}
\end{wrapfigure}

\begin{wrapfigure}[37]{R}{95mm}
  \includegraphics[width=90mm]{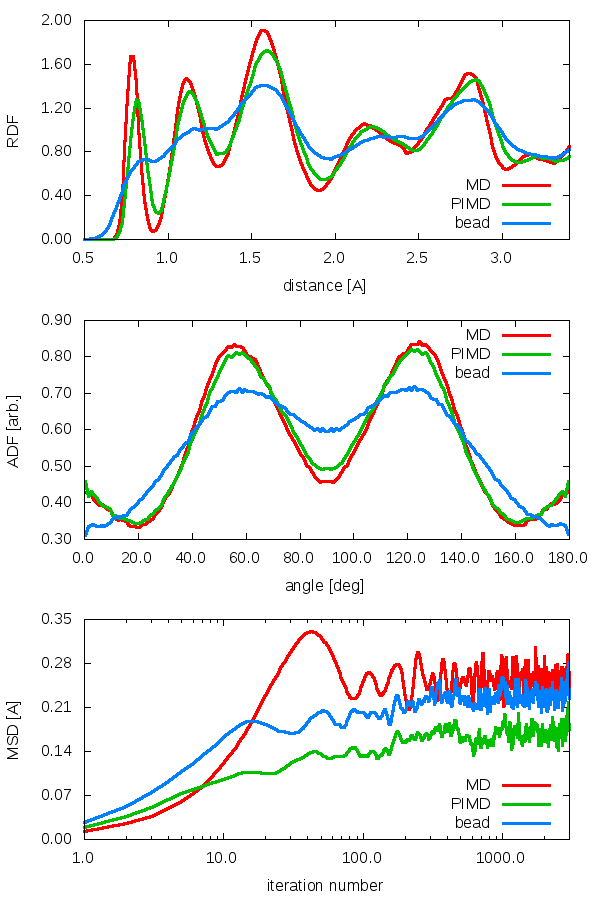}
\caption{The radial distribution function (RDF) (top), angular distribution function (ADF) (middle) and mean square-root displacement (MSD) (bottom) for ab initio MD (red), 16-bead PIMD centroid dynamics (green) and PIMD individual bead (blue).  Calculations are done for
128 atoms of $Cmca4$ hydrogen run at 400GPa and 300K, using PBE. \label{rdf}}
\end{wrapfigure}

Figure \ref{quasi} shows a number of interesting features.  The lack
of high frequency modes in $I4/amd$ reduces its zero point energy
enormously in comparison with $Cmca4$(high). The $Cmca4$(low) structure
has an even-higher branch of vibrons around 3400cm$^{-1}$, reflecting
its more molecular nature. First/second neighbours are at
0.80/1.05\AA and 0.88/0.97\AA in the two $Cmca4$ structures (from
PBESOL).   The consequent 10meV/atom difference in ZPE would be absent in AIMD:
enough to shift the transition pressure between these phases by tens of GPa.

At 300K for $Cmca4/I4amd$ phases, the quantum heat capacities are
typically 0.3R, compared with 1.5R classically.  It implies that the
vibrational entropy may be overestimated by an order of magnitude in
AIMD, which will induce a significant error on phase transition
temperatures to high entropy phases.  This may be mitigated by the
fact that for Phase IV and the melt a large contribution to the
entropy is configurational.  In fact, high entropy phases typically have lower
zero-point energy, which is an even larger effect than the quantum reduction
in heat capacity:  Reduced heat capacity favours low entropy phases,
so there is some cancellation of errors.

Comparison of the two upper panels in Figure \ref{quasi}
shows that molecular vibration frequencies are underestimated in
lattice dynamics, compared with MD, and what has been seen
experimently \cite{magduau2013identification}.  So even when ZPE is
included via lattice dynamics, its effect may be underestimated.  The
lowest panel illustrates the size of the effects.  At room
temperature, ZPE far exceeds the classical kinetic energy, yet the
quantum kinetic energy is negligible and visible only in the low
energy modes. The three different terms (ZPE, classical and quantum
KE) integrate to 327, 77 and 4 meV/atom respectively.  As well as the
effects of zero-point energy, zero-point {\it motion} is around
$0.2\AA$.  If neglected in AIMD, additional thermal motion would
required to initiate diffusion, meaning that the onset of diffusion in
the simulation is of order 100K greater than in
reality\cite{belonoshko2013atomic}.

Since these quantum effects are missing in previous MD
simulations\cite{magdau2014high,belonoshko2013atomic}, the observation of a wide field of
$Cmca4$ stability may be suspect, perhaps even sampling the wrong
$Cmca4$. 

We note a practical problem in calculating $g(\omega)$ - the phonons are
weakly coupled, and even if the perturbations are strong enough to
satisfy the KAM theorem \cite{kolmogorov1954preservation}, it may take many nanoseconds before
equipartition occurs.  Most serious of these is the overpopulation of
the ``ringing mode'' described above,
which manifests as a long term oscillation in volume (NPT ensemble) or
pressure (NVE).  
The normalization of the expression for the velocity
autocorrelation function ${\langle v(0)^2\rangle}$ assumes
equipartition.  The unresolved  dichotomy here is that we {\it assume} harmonicity
of the quantum mechanics, and {\it require} anharmonicity to achieve
equipartition.

\subsection{Non-harmonic modes: Rotons}
In  Phase I, two degrees of freedom
per molecule involve free rotations: ``rotons''.  In this case, the
energy is quantized according to $J(J+1)\hbar^2/2I$, with $J$ a quantum
number and $I$ the moment of inertia, as is readily observable in
spectroscopy\cite{howie2012mixed,magduau2014local}. A curious feature of the roton energy is that it bears no
relation to any frequency which might be measured in MD. 

Rotons have no zero point energy (J=0), and for distinguishable atoms
are described by a partition function:
\begin{gather}
Z = \sum^{\infty}_{J=0}  (2J+1) {\exp(-J(J+1)\hbar^2/2Ik_BT)}  
\end{gather}
\noindent where $2J+1$ is the degeneracy of each mode.  As with the harmonic
oscillator, the energy in high $T$ limit is $k_BT$ per mode, but
significantly lower at low T.  For ortho- and para- hydrogen the
degeneracies are different, but the limiting cases are the same.  The
energy of the $J=1$ state is of order 15meV, so it is thermally
excited at room temperature and the quantum energy approaches the
classical value.

It is also noteworthy that the ground state $J=0$ wavefunction or the
classically tumbling dimer is spherically symmetric, which helps to
explain why  the
favoured structure coresponds to the close packing of spheres.

\subsection{Zero point pressure}

As a material is compressed, the zero point energy increases.  This is a
contribution to the pressure required to compress the material.  It
can be readily calculated from ab initio lattice dynamics
calculations,  and is typically a few tens of GPa.

\begin{wrapfigure}[38]{H}{90mm}
  \includegraphics[width=90mm]{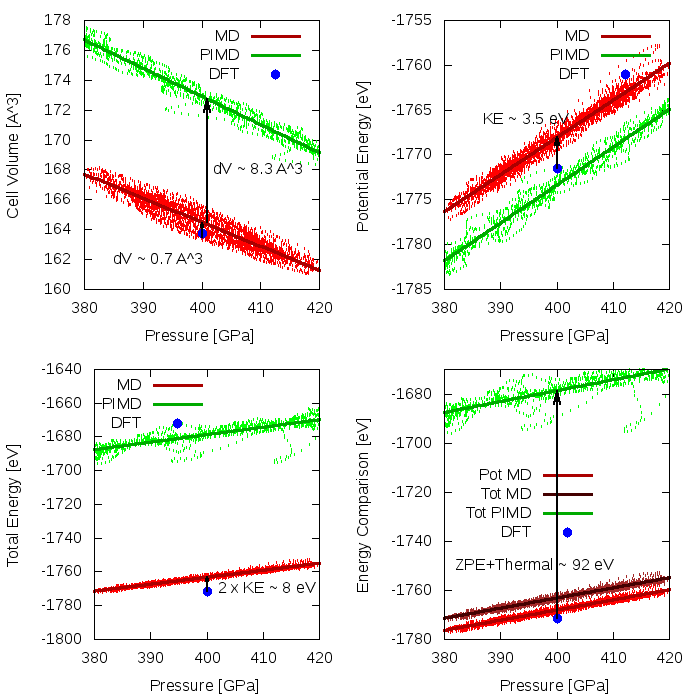}
\caption{Results from MD and PIMD runs in NPT ensemble.  top panels
  and bottom right panel compare the cell volume, potential energy and
  total energy, respectively as extracted from MD, PIMD (for all 128
  atoms, at 300 K) and single-point DFT calculation (at 0 K, with
  arrows showing thermal or ZPE shifts). The total energy in DFT is
  the same as the potential energy and is only repeated for reference,
  whereas the total energy in MD is the sum of the DFT energy and
  vibrational energy (i.e. twice the kinetic energy $(3N-3) k_BT$ in
  the classical harmonic limit). The last panel illustrates the
  zero-point plus thermal energy calculated from PIMD.  The barostat
  is weak, and we note that all quantities vary with pressure.
\label{energy}}
\end{wrapfigure}

This effect is completely absent in AIMD calculations. The zero point
pressure can be estimated from Table 1 as the ratio of $\Delta U_{ZPE}$ to ${\Delta V}$ e.g. in $I41/amd$ with LDA this is 12.5GPa. In
Figure \ref{energy} we see that ZPE pressure is enough to change the
volume of $Cmca4$ by 5\%.

It is interesting to note that the zero-point energy in a liquid must
be similar to that in a solid, since typical ZPE values around
0.25eV/atom far exceed the thermal energy at the melting temperature.

\subsection{Kinetic pressure}
Static calculations of energy minimization ignore the pressure effects
from the kinetic energy which give rise to thermal expansion.  In classical MD, this can be easily
estimated from the ideal gas law P=nRT/V.  For hydrogen at around
300GPa, we have a density of about one molecule per 2.5$\AA^3$, which
gives a classical kinetic pressure of around 1.5GPa at room
temperature.  The quantum kinetic pressure is much lower for hydrogen
at room temperature, since most modes are not excited.

\section{Conclusions}

We have shown that there is a large degree of uncertainty around the
use of molecular dynamics in the study of high pressure hydrogen.
Various treatments of the exchange-correlation functional give
somewhat differing results, accounting for tens of GPa uncertainty in
phase boundaries. The problems with finite size and definition of a
phase introduce unquantifiable errors.  The major uncertainty comes
from understanding the quantum behaviour of the protons.  In
particular, this relates to the zero point vibrations.  These
contribute massively to the total energy, and thus also introduce a
correction of tens of GPa to the nominal pressure calculated from the
Kohn-Sham energy and virial. The energy difference
between two phases can be calculated by integrating the forces along a
transition path: therefore in addition to a missing zero-point energy,
there are missing zero-point forces and pressures in molecular dynamics.

Our results also suggests that previous work with classical MD,
lattice dynamics and the PBE functionals is likely to have
systematically overestimated the stability of this molecular phases
relative to low-ZPE structure such as $I41amd$ or its high
temperature variants such as chainlike structures.

In view of all this, one might ask whether there is any point to doing
molecular dynamics.  The answer remains a resounding yes, with
appropriate modesty about numerical results. 

For perturbative effects such as vibrational
frequencies MD provides the best way to tackle anharmonicity.  Even
when qualitative accuracy is missing, MD can give indications of
relative densities for phase boundary slopes. In the NPH ensemble MD
can transform between crystal phases giving candidate structures for
comparison with experiment.  Calculated barrier heights can be used to
determine when tunneling effects should augment thermal ones in
considering diffusion.  MD can give snapshot indications of the
structure of the melt.  Without MD the character of Phase IV would be
unknown\cite{liu2013proton}, the liquid structure with its high
density would be a mystery and the discrepancy between lattice
dynamics and experimental Raman frequency would remain unsolved.

 Many material properties
remain stubbornly inaccessible to experiment.  If density could have
been measured as reliably as pressure, Wigner and Huntingdon's
prediction\cite{wigner1935possibility} would be regarded as correct:
to avoid their much derided 25GPa figure, they needed to know that the
bulk modulus was ten times that of steel, while the thermal expansion
remains similar.  AIMD shows that the interesting behaviour occurs
when energy per atom is around 13.5eV/atom (Fig. \ref{energy}), very
close to the energy of a hydrogen atom, i.e. when the total binding
energy is close to zero and the atoms are primarily kept together by external pressure.

\section*{References}

\bibliographystyle{iopart-num}
\bibliography{Refs}

\newpage

\end{document}